\documentstyle [twocolumn,amssymb,epsf,psfig]{mn}
\newcommand{\mincir}{\raise
-3.truept\hbox{\rlap{\hbox{$\sim$}}\raise4.truept\hbox{$<$}\ }}
\newcommand{\magcir}{\raise
-3.truept\hbox{\rlap{\hbox{$\sim$}}\raise4.truept\hbox{$>$}\ }}
\newcommand{\minmag}{\raise
-3.truept\hbox{\rlap{\hbox{$<$}}\raise5.truept\hbox{$<$}\ }}
\newcommand{\be}{\begin{equation}}
\newcommand{\ee}{\end{equation}}
\newcommand{\ba}{\begin{eqnarray}}
\newcommand{\ea}{\end{eqnarray}}
\newcommand{\brr}{\begin{array}}
\newcommand{\err}{\end{array}}
\newcommand{\bc}{\begin{center}}
\newcommand{\ec}{\end{center}}

\title[Serendipitous X-ray cluster survey with {\it XMM-Newton}]     
{The Serendipitous XMM-Newton Cluster Athens Survey (SEXCLAS): Sample 
selection and the cluster {\boldmath $\log N - \log S$}}     
\author[Kolokotronis et al.]
{V. Kolokotronis$^{1}$, A. Georgakakis$^{1}$, S. Basilakos$^{1}$,
I. Georgantopoulos$^{1}$, 
\\ \\
{\LARGE M. Plionis$^{1,2}$, S. Kitsionas$^{1,4}$, T. Gaga$^{1,3}$}
\\ \\
$^1$ Institute of Astronomy \& Astrophysics, National Observatory of Athens, 
I.Metaxa \& B.Pavlou, Palaia Penteli, 152 36, Athens, Greece \\
$^{2}$ Instituto Nacional de Astrofisica, Optica y Electr\'onica (INAOE)
Apartado Postal 51 y 216, 72000, Puebla, Pue., Mexico \\
$^3$ Physics Department, Univ. of Athens, Panepistimioupolis,
Zografou, Athens, Greece \\
$^4$ Astrophysikalisches Institut Potsdam, An der Sternwarte 16, D-14482, 
Potsdam, Germany \\
}

\begin{document}

\maketitle

\begin{abstract}

\indent In this paper we serendipitously identify X-ray cluster candidates 
using {\it XMM-Newton} archival observations complemented by 5-band
optical photometric follow-up observations ($r\approx23$\,mag) as part of 
the X-ray Identification (XID) programme. Our sample covers an area of 
$\rm\approx 2.1\,deg^2$ (15 {\it XMM-Newton} fields) and comprises a total 
of 21 (19 serendipitous + 2 target) extended X-ray sources to the limit 
$f_{\rm x}\,(\rm 0.5 - 2\,keV)\approx 6\times 10^{-15}\,erg\,s^{-1}\,cm^{-2}$, 
with a high probability ($ > $99.9 \%) of being extended on the 
{\it XMM-Newton} images. Of the 21 cluster candidates 7 are spectroscopically 
confirmed in the literature. Exploiting the optical data available for these 
fields we discover that $\ga 68 \%$ of the X-ray cluster candidates are 
associated with optical galaxy overdensities. We also attempt to constrain 
the redshifts of our cluster candidates using photometric methods. We thus 
construct the photometric redshift distribution of galaxies in the vicinity 
of each X-ray selected cluster candidate and search for statistically 
significant redshift peaks against that of the background distribution of 
field galaxies. Comparison of the photometric with spectroscopic redshift 
estimates for the confirmed clusters suggest that our simple method is robust 
out to $z\approx 0.5$. For clusters at higher-$z$, deeper optical data are 
required to estimate reliable photometric redshifts. Finally, using the 
sample of the 19 serendipitous X-ray selected cluster candidates we estimate 
their surface density down to $f_{\rm x}\,(\rm 0.5 - 2\,keV)\approx 
6\times 10^{-15}\,erg\,s^{-1}\,cm^{-2}$ and find it to be in fair agreement 
with previous and recent studies.
\end{abstract}

\begin{keywords}
Surveys: galaxies: clusters; Cosmology: large--scale structure of Universe; 
Surveys 
\end{keywords}

\section{Introduction}\label{introduction}

Since the first detection of the Virgo and Coma clusters at X-ray
wavelengths (Byram, Chubb \& Friedman 1966; Meekins et al. 1971), a
large observational effort has been put forward aiming to compile
X-ray cluster samples over a wide range of redshifts and
luminosities. Such programs are mainly driven by the realisation that
galaxy clusters, the most massive virialised systems known, are
prime diagnostic tools for both cosmological models and structure
formation theories (eg. Bahcall  1988; Borgani \& Guzzo 2001; Rosati,
Borgani \& Norman 2002). 
 
Eventhough the first all-sky cluster catalogues were carried out at
optical wavelengths (Abell 1958; Zwicky et al.  1968; Abell, Corwin \&
Olowin 1989; Lumsden et al. 1992; Dalton et al. 1994), problems related to 
projection effects and complex selection criteria led to the search for 
alternative methods of compiling cluster samples. In this respect the X-ray 
wavelengths offer a crucial feature: the X-ray emission of galaxy clusters is 
due to centrally concentrated hot gas that is relatively easy to identify 
against the X-ray sky. Therefore, X-ray selected samples are less prone to 
projection biases corrupting optical catalogues.

Several X--ray cluster samples have been accumulated and used for
a variety of astrophysical and cosmological studies. The first
all-sky X--ray cluster sample observed by {\it UHURU} contained 52
entries (Forman  et al. 1978). This was followed by further X-ray
missions, {\it Ariel-V} and {\it HEAO-1} that achieved even deeper
observations (Cooke et al. 1978;  Piccinotti et al. 1982). The launch
of the {\it Einstein} Observatory, the  first with imaging capabilities, 
provided a step forward in X-ray cluster astronomy. The {\it Einstein} 
Medium Sensitivity Survey covers almost $1000$\,deg$^2$ yielding a 
flux-limited sample of $\ga 100$ clusters (Gioia et al. 1990; 
Henry et al. 1992). The advent of the {\it ROSAT} mission, with 
improved sensitivity  and spatial resolution offered a further 
significant boost to cluster studies, providing samples over a wide 
range of depth, redshift and luminosity (Romer et al. 1994; Ebeling et 
al. 1996; 1998; 2001; Rosati et  al. 1998; Vikhlinin et al. 1998; 
Romer et al. 2000; B\"{o}hringer et al.  2000; 2004; Perlman et al. 2002).  

Recently, the {\it XMM-Newton} with 10 times more effective area and
5 times better spatial resolution than {\it ROSAT} provides an ideal
platform to study clusters out to high-$z$ (Pierre et al. 2004;
Valtchanov et al. 2004). In addition to observational 
programs specifically designed to compile cluster samples (Pierre et al. 
2004), the huge {\it XMM-Newton} public database also provides opportunities
to perform serendipitous cluster surveys (Romer et al. 2001; Lamer et al. 
2003; Basilakos et al. 2004 hereafter BPG04; Gaga et al. 2005 hereafter 
GPB05; Mullis et al. 2005). Land  et al. (2005; hereafter LND05) for example, 
have combined public {\it XMM-Newton} observations with the Sloan Digital 
Sky Survey (SDSS; Abazajian 2003) and found that many of their extended 
X-ray sources are not associated with SDSS galaxy overdensities, indicating 
either high-$z$ systems that require deeper optical data ($r > 22.5$) or 
groups of galaxies not standing out against the background. Furthermore, 
Plionis et al. (2005) using public XMM-{\em Newton} observations studied the 
X-ray properties of a subset of the Goto et al. (2002) optical SDSS clusters. 
They found that less than half of their 17 optically selected clusters have 
X-ray emission with a flux $f_{\rm x}\,(\rm 0.5-2\,keV)\;\magcir 1.2\times 
10^{-14}$ erg cm$^{-2}$ s$^{-1}$. The remaining SDSS clusters have a 
3$\sigma$ upper limit corresponding to $L_{\rm x}\;\mincir 5\times 10^{42}$ 
erg/sec, implying very poor systems if real at all.

This paper presents first results from the ongoing Serendipitous X--ray 
Cluster Athens Survey (SEXCLAS) using public {\it XMM-Newton} observations 
supplemented by 5-band optical photometry from the INT {\it XMM-Netwon} 
Serendipitous Source Catalogue (SSC) XID program. In this first paper, we 
describe the selection of X-ray clusters, explore their association with 
optical galaxy overdensities, compare photometric with spectroscopic redshifts 
when possible and finally estimate their surface density. Compared 
to the LND05 study, our survey has somewhat deeper optical observations 
($r\approx 23$) providing an advantage when studying the association of 
X-ray clusters with optical galaxy  overdensities. Additionally, we go a 
step further and estimate the $\log N - \log S$ of X-ray selected clusters. 
In what follows we employ a flat cosmology with $\Omega_{\Lambda}=0.7$ and 
$H_{\circ}=100\,h$ km s$^{-1}$ Mpc$^{-1}$.  

\begin{table*}
\caption[]{Observing log of our survey. Columns are as follows: 1: index 
number, 2: XMM-{\em Newton} field name, 3, 4: equatorial coordinates
of field center, 5:  galactic latitude $b$ (in degrees), 6: good time 
interval in ks.}
\label{tab1}
\tabcolsep 14pt
\begin{tabular}{clcccc} 
\hline
 Index  & Field &    $\alpha$ & $\delta$&  $b$  & PN exp. time \\  
 number & name  &     (J2000) & (J2000) & (deg) & (ks)         \\      
 \hline 
 1  & CL 0016+16    & 00 18 40.3 & $+$16 27 39.7 & $-$45.71   & 25.6 \\
 2  & Mrk 1014      & 01 59 43.5 & $+$00 22 21.4 & $-$57.94   & 4.8 \\
 3  & SDS-1         & 02 17 53.5 & $-$05 01 17.7 & $-$59.75   & 40.2 \\
 4  & SDS-2         & 02 19 29.6 & $-$05 01 06.5 & $-$59.49   & 37.8 \\
 5  & GL 182        & 04 59 26.4 & $+$01 46 04.9 & $-$23.76   & 15.7 \\
 6  & MS 0737.9+7441& 07 44 34.8 & $+$74 34 12.5 & $+$29.57   & 27.0 \\
 7  & PG 0844+349   & 08 47 31.5 & $+$34 44 46.3 & $+$37.96   & 7.3 \\
 8  & Lockman Hole  & 10 52 29.1 & $+$57 29 36.9 & $+$53.14   & 33.7 \\
 9  & MS 1137.5+6625& 11 40 11.1 & $+$66 10 24.2 & $+$49.45   & 13.0 \\
 10 & Mkn 205       & 12 22 10.1 & $+$75 17 28.0 & $+$41.67   & 10.7 \\
 11 & HD 117555     & 13 30 38.2 & $+$24 13 51.1 & $+$80.68   & 33.0 \\
 12 & PKS 2126-158  & 21 29 04.6 & $-$15 39 59.9 & $-$41.87   & 6.1 \\
 13 & PKS 2135-147  & 21 37 38.5 & $-$14 34 23.9 & $-$43.33   & 30.0 \\
 14 & IRAS 22491-18 & 22 51 42.6 & $-$17 53 52.4 & $-$60.95   & 15.9 \\
 15 & EQ Peg        & 23 31 43.9 & $+$19 54 43.1 & $-$39.14   & 9.3 \\
\hline
\end{tabular}
\end{table*}

\section{Observations}

\subsection{Field selection}

In this paper we use public {\it XMM-Newton} data with follow-up 
multiwaveband optical observations available as part of the {\it XMM-Netwon} 
SSC XID programme. From a total of 77 {\it XMM-Newton} fields with optical 
photometric data available, we select those that (i) have imaging in at least 
5 bands ($U,g,r,i,z$ filters) to allow multicolor study of the cluster member 
galaxies, (ii) lie at high Galactic latitude $|b|>20^{\circ}$ to avoid high 
hydrogen Galactic column densities and  contamination by Galactic stars and 
(iii)  have been observed by {\it XMM-Netwon} with the EPIC-PN in full-frame 
mode. A total of 15 fields fulfill these criteria, 2 of which have clusters of 
galaxies as prime targets. Details on individual observations are presented 
in Table \ref{tab1}. Note that cluster targets are in fields \#1 and \#9.

\subsection{X-ray data}

The {\it XMM-Newton} data have been analysed using the Science
Analysis Software (SAS 5.4.1). Event files for the PN and the two MOS
detectors have been produced using  the {\sc epchain} and {\sc
emchain} tasks of SAS respectively. The event files were screened for
high particle  background periods by rejecting times with 0.5-10\,keV
count rates higher than 25 and 15\,cts/s for the PN and the two MOS
cameras respectively. The PN good time intervals for the fields used in this 
paper are shown in Table \ref{tab1}. Images in celestial coordinates 
with pixel size of 4.35$^{''}$ have been extracted in the 0.5-2\,keV spectral 
band for both the PN and the MOS event files. Exposure maps accounting for 
vignetting, CCD gaps and bad pixels have also been constructed. 

\subsection{Optical data}\label{opti} 

The optical data for this project have been obtained at the 2.5\,m INT 
telescope using the Wide Field Camera (WFC) as part of the XMM-{\em Netwon} 
SSC XID programme\footnote{http://xmmssc-www.star.le.ac.uk/}. The 
WFC is mounted at the prime focus of the INT and comprises 4 thinned EEV 
4kx2k CCDs with a pixel scale of 0.33$^{''}$. The total sky coverage per 
exposure is $\rm 0.29\,deg^2$. The multiwaveband observations ($U,g,r,i,z$ 
filters) are reduced using the pipeline reduction of the CASU INT Wide Field 
Survey\footnote{www.ast.cam.ac.uk/$\sim$wfcsur/}, resulting in 
photometrically and astrometrically calibrated images. The exposure times 
are typically 10\,min for the $g,r$ filters and 40\,min for $U,i,z$ filters.

\begin{figure} 
\mbox{\epsfxsize=8.21cm \epsffile{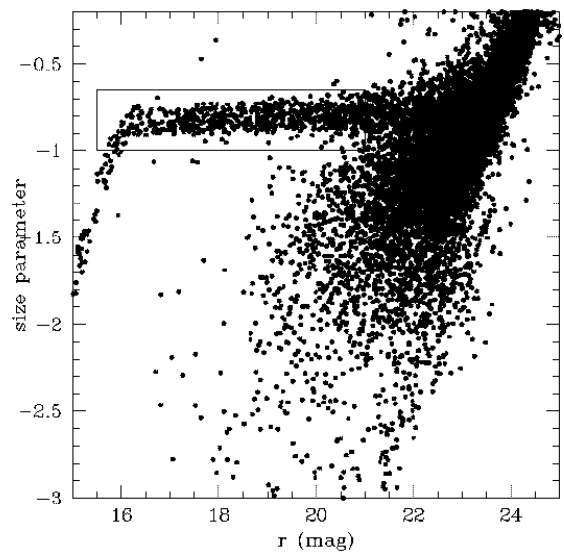}}
\caption{Star-galaxy separation diagram for the $r$-band data taken from 
field \#1 of Table 1. The size parameter is defined as the difference between 
$m_{\rm c}$ and $m_{\rm t}$ (see text for details). The stellar sequence is 
demarcated by the solid line box.}
\label{fig_star_gal} 
\end{figure}

\begin{figure}
\mbox{\epsfxsize=8.21cm \epsffile{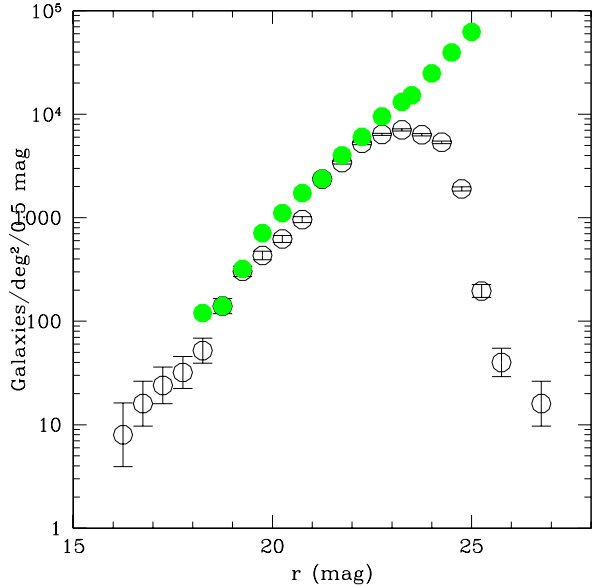}}
\caption{$r$-band galaxy counts of a typical INT observation (same as in 
Figure \ref{fig_star_gal}) used in this paper (open circles). The filled 
circles are the galaxy counts of Metcalfe et al. (1991).}
\label{fig_rcounts}  
\end{figure}

Source extraction and photometry is performed using the SExtractor
package (Bertin \& Arnouts 1996) with parameters  (detection threshold
and minimum area for detection) tuned to minimise the number of
spurious detections, while ensuring that faint sources are included in
the final catalogue. Regions contaminated by bright stars or bad pixels
are masked during the source extraction. For the star-galaxy separation
we have considered a size parameter, defined as the difference between the
core, $m_{\rm c}$, and `total', $m_{\rm t}$ magnitude of sources in the 
$r$-band . The former corresponds to the intensity  within an aperture with 
size similar to that of the seeing at the time of the observation and the 
latter is the Kron magnitude estimated by the SExtractor. As an example, 
Figure \ref{fig_star_gal} plots the difference between the `total' and 
`core' magnitude (size parameter) against the `total' magnitude in the case 
of a typical INT observation in the $r$-band (field \#1 of Table 1). The 
stellar sequence is demarcated with a solid lined rectangular box. The 
distribution of stars and galaxies in Figure \ref{fig_star_gal} overlaps at 
$r\ge\,$21\,mag. At fainter magnitudes no attempt is made to further 
eliminate stars from the sample, since compact galaxies could be mistakenly 
removed. Furthermore, the number of stars relative to galaxies becomes 
increasingly smaller beyond this magnitude. The $r$-band galaxy counts of a 
typical INT pointing (same as in Figure \ref{fig_star_gal}) are shown in 
Figure \ref{fig_rcounts} along with the compilation of Metcalfe et al. (1991). 
At $r\le\,$23\,mag, our results are in good agreement with the Metcalfe et 
al. (1991) number counts. At fainter magnitudes our sample is affected by 
incompleteness.

Moreover, we exploit the 5-band optical photometry in order to estimate 
photometric redshifts ($z_{\rm p}$) for our sources using the {\sc hyper-z} 
code (Bolzonella, Miralles \& Pell\'o 2000). The {\sc hyper-z} program 
determines the $z_{\rm p}$ of a given object by fitting a set of template 
Spectral Energy Distributions (SEDs) to the observed photometric data 
through a standard $\chi^2$ minimisation technique. The template rest-frame 
SEDs used here are the observed  mean spectra for four different galaxy 
types (E/S0, Sbc, Scd, Im) from Coleman, Wu \& Weedman (1980) extended in the 
UV and IR regions using the spectral synthesis models of Bruzual \& Charlot 
(1993) with parameters selected to match the observed spectra. 

Photometric redshifts are estimated only for those sources with at least 
4-band (the 4 redder) photometric information available. Comparison with 
spectroscopic redshifts found in the literature (NED) enables us to 
approximate the accuracy of our $z_{\rm p}$ which is thus computed via 
$\delta z_{\rm p}/(1+z_{\rm p})\approx 0.1$ and holds true for 
$z_{\rm p}\le 0.6$. This appears as an upper limit to the reliability of 
the estimated photometric redshifts of our X--ray selected clusters, which 
can be explained by the different limiting magnitudes of the 5 bands taken 
into account in the $z_{\rm p}$ estimates.
  
\section{X-ray cluster selection}\label{xray}

The source detection is performed on the 0.5-2\,keV PN image using the 
{\sc ewavelet} task of SAS with a $5\sigma$ detection threshold. We 
use the PN image due to the higher sensitivity of the EPIC-PN compared 
to the MOS detectors. A by-product of the source detection algorithm 
are smooth background maps providing an estimate of the background 
counts at each position. Cluster candidates are identified by 
searching for X-ray extended sources using the {\sc emldetect} task of 
SAS. This uses the {\sc ewavelet} source list as input and performs 
multi-PSF maximum likelihood fits to the count distribution of 
individual sources to assign a probability that an object is 
extended. We select clusters with {\sc emldetect} extension 
probability $> 99.9 \%$. Visual inspection reveals that the 
choice of probability cutoff ensures that all obvious clusters are 
included in the catalogue while minimising the number of spurious 
detections. After excluding a total of 17 sources that clearly lie on 
CCD gaps or are related to double point sources, we finally extract a total 
of 21 cluster candidates with the above extension probability. These are 
presented in Table 2 together with their X--ray and optical properties. 

\begin{figure*}
\mbox{\epsfxsize=8.21cm \epsffile{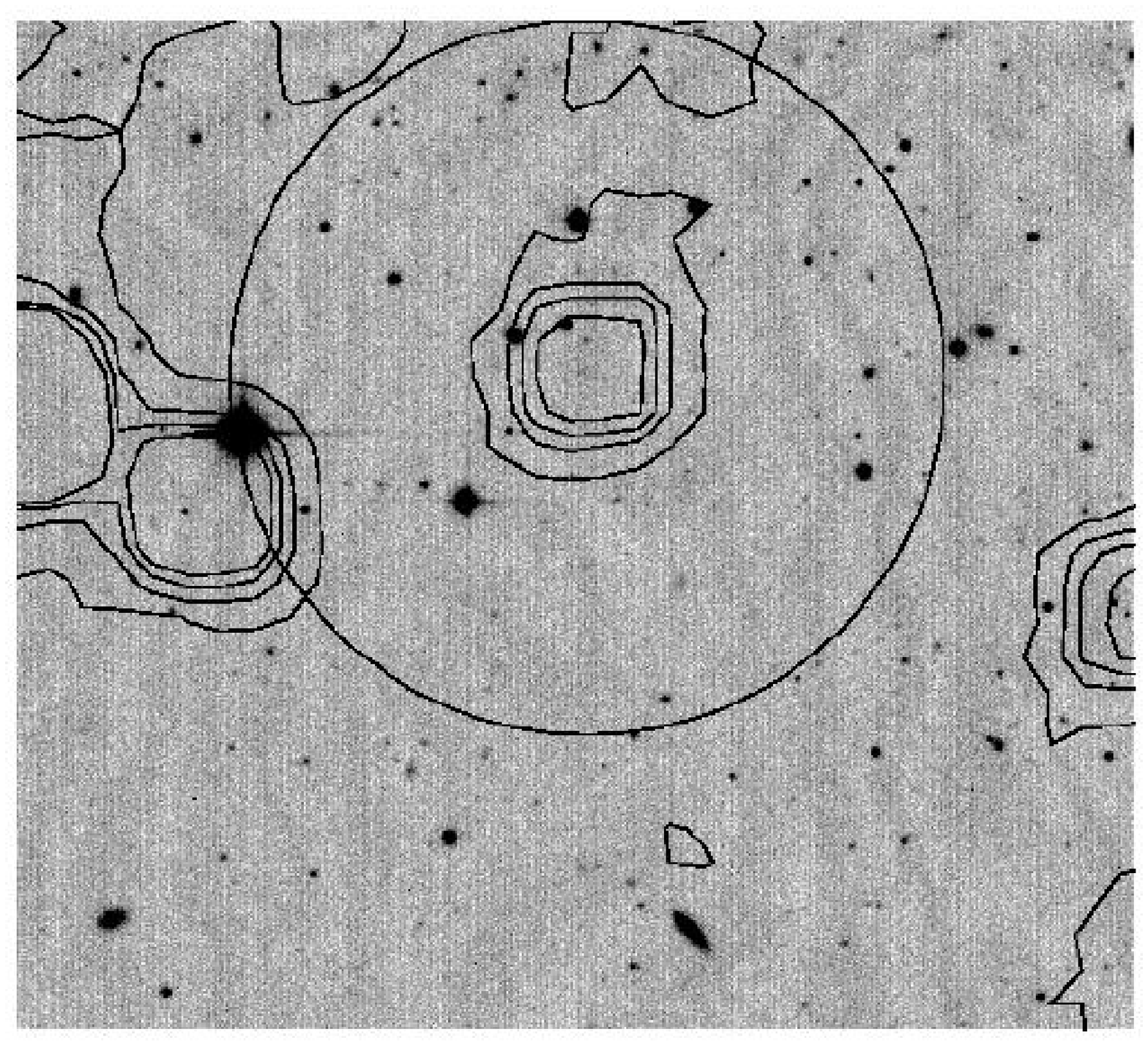}}
\mbox{\epsfxsize=7.60cm \epsffile{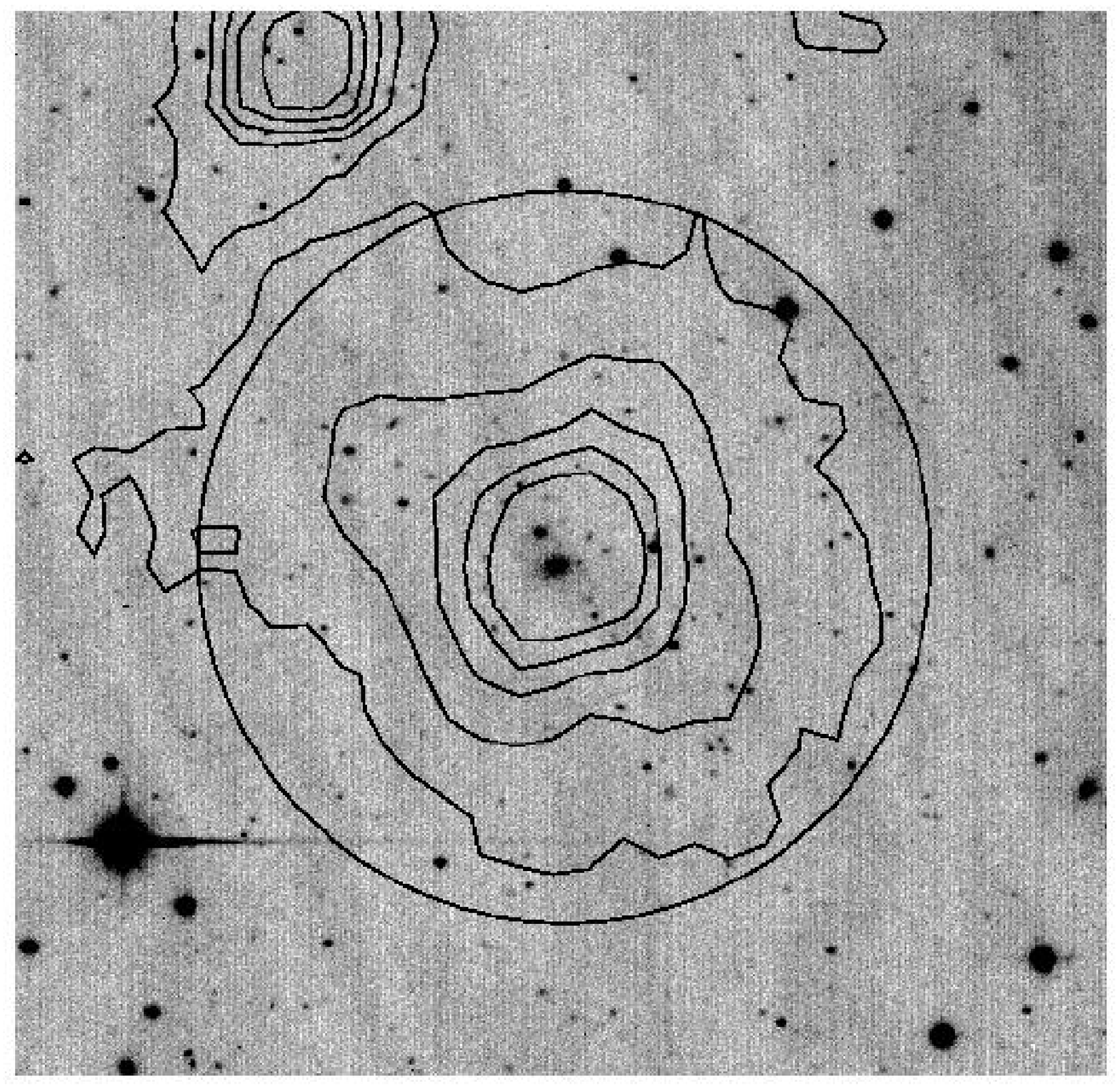}}
\caption{Underlying optical images versus X--ray gas contours for two 
distinct cases. {\bf Left panel}: Distant cluster HST J001831+16207 
(Ostrander et al. 1998) which appears as a point-like X-ray source in our 
survey. {\bf Right panel}: A spectroscopically confirmed X--ray cluster 
associated with a nearby significant optical overdensity (object \#1 of 
Table 2). In both panels, circles mark a $3^{'}$ diameter.}
\label{xopti} 
\end{figure*}

Spectroscopic redshifts are available for 7 out of the 21 cluster candidates,  
2 of which are the prime targets of the respective XMM pointings (objects \#2 
and \#16 in Table 2). For the remaining 14 X-ray selected cluster candidates 
there is no spectroscopic information available in the literature. The most 
distant cluster is at $z\sim 1.13$ (RX\,J1053.7+5735; Hashimoto et al. 2005), 
while the nearest lies at a $z = 0.386$ (VMF\,98-021). A typical image of a 
spectroscopically identified X--ray cluster clearly associated with a nearby 
significant optical overdensity is shown in the right panel of 
Figure \ref{xopti} (object \#1 of Table 2).

For the flux estimation we use a circle with radius in the range
$18^{''}$ to $30^{''}$ depending on the extent of the cluster on the 
{\it XMM-Newton} images. Count rates are converted to fluxes by choosing a 
Raymond-Smith model SED with temperature $T \approx 2$\,keV and Galactic 
absorption appropriate for each field. We also apply a correction to the 
estimated fluxes to account for the cluster emission outside the aperture 
used to sum the source counts. We adopt a King's surface brightness profile 
with core radius $r_{\rm c}=0.1\,h^{-1}$\,Mpc (Rosati et al. 1995; 1998) to 
estimate the flux fraction outside the aperture used. We convert radial 
apertures to physical coordinates using either the spectroscopic redshift 
($z_{\rm s}$) of the cluster (available for 7 systems) or $z_{\rm p}$ 
estimates described in section \ref{photoz}. In some cases there is no 
$z_{\rm s}$ or $z_{\rm p}$ and we assume $z=0.4$, which coincides with the 
median expected redshift of the {\it ROSAT} deep cluster survey (Rosati et 
al. 1998). The corrected fluxes are presented in column 5 of Table \ref{tab2}.

We note that the {\sc emldetect} extended source identification
algorithm produces reliable results when there is sufficient
signal--to--noise ratio to perform multi-PSF maximum likelihood
fits. It is therefore possible that high-$z$ or intrinsically faint
clusters with few photons may appear point-like on the {\it
XMM-Newton} images and therefore missed from our sample. For example,
Ostrander et al. (1998) searched for optical galaxy overdensities in
the Hubble Space Telescope Medium Deep Survey and compiled a sample of
optical cluster candidates. One of their systems, HST\,J001831+16207 
(their Figure 5), overlaps with our survey (field \#1) and coincides with 
an X-ray detected source classified as point-like by the {\sc emldetect} 
algorithm (see left panel of Figure \ref{xopti}). Limited spectroscopic 
information suggests an overdensity of optical galaxies at $z\approx 1.3$ 
(Yan \& Thompson 2003; Yan et al. 2004) providing evidence that this might 
indeed be a real cluster.   

\section{Optical identification}

In this section we exploit the 5-band optical photometry available for the 
surveyed area so as to study the optical properties of the X-ray selected 
cluster candidates. This is to search for optical galaxy overdensities, 
assess the reliability of our technique at least on spectroscopically 
identified systems (cf. right panel of Figure \ref{xopti}) and provide 
redshift estimates for those where spectroscopy is not available.

\subsection{Galaxy overdensities}

We identify optical galaxy overdensities close to an X-ray selected
cluster candidate using the smoothing and percolation technique described 
in detail by BPG04. Here we only briefly discuss the most salient details 
of the method. We smooth the projected galaxy population using a Gaussian 
kernel to produce continuous density maps. Following BPG04, we 
adopt a Gaussian kernel radius of 28.5$^{''}$ corresponding to 
$\sim 0.2\,h^{-1}\,$Mpc and a cell size of about 19$^{''}$ corresponding 
to $\sim 0.15\,h^{-1}\,$Mpc both at $z=0.4$, the mean $z_{\rm p}$ of our 
optical data which is also comparable to the average redshift of the SDSS data 
reaching a similar but somewhat shallower depth (cf. Blanton et al. 2003). 
Clusters are identified on the resulting maps by searching for peaks above a 
projected overdensity of $\delta > 1$ and sizes larger than about 
0.3$\,h^{-1}\,$Mpc. 

We exclude from the optical overdensity analysis objects \#20 and \#21 in 
Table \ref{tab2} that lie within masked areas and hence have no optical data 
available. For the remaining cluster candidates, we apply the BPG04 technique 
to the optical data and identify projected optical galaxy overdensities 
in the vicinity of 13 out of 19 ($\ga 68$ \%) X-ray clusters, a finding also 
in broad agreement with recent analyses (Donahue et al. 2002; BPG04; 
Plionis et al. 2005). For the remaining 6 X-ray clusters, we do not find 
statistically significant optical overdensities insinuating either 
high-$z$ systems that need deeper optical data or poor groups with few 
members. A further possibility could be that projection effects smear out 
possible optical overdensities. However, these may appear in the photometric 
distribution (object \#13 in Table 2). Cluster candidates coupled with 
optical galaxy overdensities are presented in column 7 of Table \ref{tab2}.

It is interesting to mention that of the 7 previously known X--ray clusters, 
the BPG04 method has picked 6, the one missing being located within a masked 
region (object \#20 of Table 2). We regard such a good performance as a 
measure of the technique's robustness. 

\begin{figure}
\mbox{\epsfxsize=14.0cm \epsffile{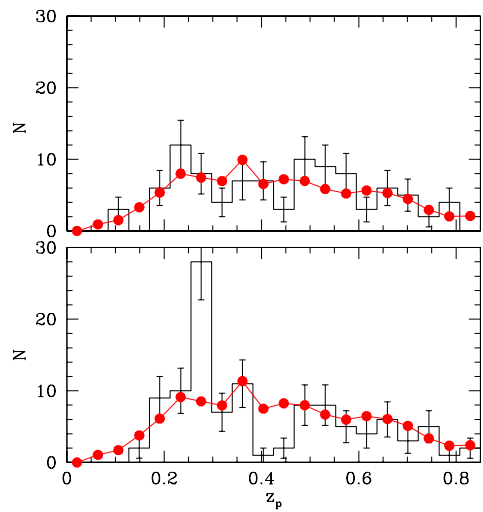}}
\caption{Photometric redshift distributions for two X--ray cluster 
candidates (histograms) over the total background $z_{\rm p}$ distribution 
(filled symbols). Poissonian errors are shown. Upper and lower panels 
correspond to clusters taken from fields \#8 (object \#14 of Table 2) and 
\#15 (object \#19 of Table 2), respectively.}
\label{xmmnz}   
\end{figure}

\subsection{Cluster photometric redshifts}\label{photoz}

\begin{table*}
\caption[]{X-ray cluster candidates. The  columns are: 1: index, 2: name 
from NED (if available), 3: right ascension $\alpha$, 4: declination 
$\delta$, 5: 0.5-2\,keV flux, 6: 0.5-2\,keV X--ray luminosity, 7: existence 
or not of an optical galaxy overdensity ($\delta_{\rm g}$), 8: Kolmogorov 
probability ${\cal P_{\rm K}}$ of significant peaks in the $z_{\rm p}$ 
distribution, 9: estimated  photometric redshift, 10: spectroscopic redshift 
(if available) and 11: notes on individual objects. The latter includes 
usually target entries, distances from NED coordinates or even masked 
regions. The first 19 objects are located in {\em clean} areas and the last 
2 fall entirely within masked areas of the optical image. Finally, entries 
with asterisks are also included in the X--ray cluster database BAX 
(http://bax.ast.obs-mip.fr/).}
\label{tab2}
\tabcolsep 4.05pt
\begin{tabular}{ccccccccccc} 
\hline
 Index & Cluster &$\alpha$ &$\delta$ &$f_{\rm x}$ &log$\,L_{\rm x}$& 
$\delta_{\rm g}$&${\cal P_{\rm K}}$ &$z_{\rm p}$& $z_{\rm s}$ & Notes \\  
 number& name  &(J2000)  &(J2000)  &($\rm 10^{-14}$ cgs) & & & & & \\ 
\hline

1 &RX J0018.2+1617$^{\ast}$&00 18 16.8&+16 17 39.8&4.05&43.69&$\surd$& 
0.056 & 0.45 & 0.55  & $\sim 8^{'}$ from \#2  \\
2 &MS 0015.9+1609$^{\ast}$&00 18 33.4&+16 26 11.4&51.24&44.78&$\surd$&
0.0038&0.45  & 0.54  &  target  \\
3& -- &02 00 19.1&+00 19 33.2&3.01 &43.48 &$\surd$& 0.01 & 0.51 & --  & -- \\
4&        --       &02 17 35.2&$-$05 13 26.0&1.17  & -- &$\surd$& 0.3  & 
-- & --   & -- \\
5&        --       &02 17 36.6&$-$04 59 24.7&0.66  & -- &$\surd$& 0.36  & 
-- & --   &     -- \\
6&        --       &02 19 34.6&$-$05 08 57.0 &2.47  & -- &$\times$& 0.483& 
--   & --   &     -- \\
7&        --       &02 19 44.6&$-$04 53 23.6&1.27  & -- &$\surd$ & 0.154& 
--   & --   &     -- \\
8&        --       &02 19 44.8&$-$04 48 39.0&0.69  & -- &$\times$& 0.404& 
--   & --   &     -- \\
9&        --       &04 59 07.3&+01 54 47.7& 2.22  & -- &$\times$&0.73  & 
--   & --   & --  \\
10&        --       &08 47 02.2&+34 51 23.4&1.27  &43.11 &$\surd$& 0.0002& 
0.51 & --   & $\sim 3^{'}$ from \#11 \\
11&VMF 98-059$^{\ast}$&08 47 10.3&+34 48 57.6&3.88  &43.69 &$\surd$&
$10^{-8}$&0.52& 0.56  & $0.37^{'}$ from cluster \\
12&        --       &10 52 38.2&+57 30 49.3&0.8   &43.10 &$\surd$ & 
0.0004&0.61 & --    & $\sim 3^{'}$ from \#13  \\
13&        --       &10 52 54.2&+57 32 09.6&0.62  &42.93 &$\times$& 0.0001 
& 0.58 & --    &    -- \\
14&RX J1053.3+5719$^{\ast}$&10 53 18.7&+57 20 38.0&2.57&43.00&$\surd$&
0.66  & -- & 0.34  & $0.15^{'}$ from cluster\\
15&RX J1053.7+5735$^{\ast}$&10 53 40.3&+57 35 24.0&1.92&44.25  &$\times$& 
0.8  & --   & 1.13 & $0.42^{'}$ from cluster\\
16&MS 1137.5+6625$^{\ast}$&11 40 23.0&+66 08 16.8&10.15&44.47 &$\surd$& 
0.1  & 0.36 & 0.782 & target  \\
17&        --       &21 36 59.5&$-$14 35 07.4&1.57  & -- &$\times$& 
0.5   & --  & --    &    -- \\
18&        --       &22 51 45.7&$-$18 05 38.4& 1.20  & -- &$\surd$& 
0.58 & -- & --   &     -- \\
19&        --     &23 32 27.4&+19 58 04.8&4.74  &43.03 &$\surd$ & 
0.0016 & 0.27 & --   &     -- \\
\hline
\hline
20&VMF 98-021$^{\ast}$&01 59 16.9&+00 30 07.2&26.7 & 44.15 &$\times$& --   & 
--  & 0.386  & mask \& cluster \\
21&        --        &08 48 16.8&+34 36 09.0&3.14 & -- &$\times$&   --   & 
--  & --    & mask \\
\hline

\end{tabular}
\end{table*}

In this section we use the $z_{\rm p}$ information in an
attempt to constrain the redshifts of the X-ray selected cluster
candidates. We construct the $z_{\rm p}$ distribution in the
vicinity of a given X-ray selected cluster by extracting all galaxies
within a $\sim 2.5^{'}$ radius around the X-ray centroid. This search radius 
corresponds to a physical separation of $r_{\rm s}\sim 0.85\,h^{-1}$\,Mpc at 
$z\sim 0.4$, the mean redshift of our photometric data. We then search for 
statistically significant peaks in the $z_{\rm p}$ distribution, implying the 
presence of a galaxy overdensity, by comparing with the mean photometric 
distribution of the galaxies in all the available fields. The background 
galaxy distribution is scaled to the number of galaxies extracted in the 
vicinity of the X-ray selected cluster.

We attempt to quantify the differences between the two distributions 
(background and generic cluster) using the Kolmogorov-Smirnov test to 
estimate the probability $\cal P_{\rm K}$ that they are drawn from the same 
parent population. A low probability may suggest the presence of galaxy 
overdensities, i.e. peaks in the distribution of galaxies in the vicinity of 
the X-ray selected cluster. We apply a cutoff in the probability 
${\cal P_{\rm K}}=0.1$ to limit the sample with $z_{\rm p}$ 
estimates only to those clusters with statistically significant peaks against 
the background. We note, however, that there are X-ray cluster candidates 
with ${\cal P_{\rm K}}>0.1$, suggesting small differences between background 
and cluster distributions, where we can still identify peaks in the 
$z_{\rm p}$ distribution at the $\approx2\sigma$ level (cf. objects \#4, \#5 
of Table 2). This may hint at a group or a poor cluster with few members 
showing against the background distribution. Two examples of $z_{\rm p}$ 
distributions are presented in Figure \ref{xmmnz}. The upper plot (object 
\#14 of Table 2) depicts a previously known X--ray cluster with 
${\cal P_{\rm K}}>0.1$, while the lower shows a prime X--ray cluster 
candidate as it is obvious from a visual look-over (object \#19 of Table 2).

For clusters with ${\cal P_{\rm K}}<0.1$, the approximate $z_{\rm p}$ of a 
peak is firstly estimated via visual examination. In the case 
of many peaks we select the most statistically significant one on the basis 
of Poisson statistics. For a more accurate redshift estimate we adopt the 
method of LND05 and apply a Gaussian fit to the distribution in 
the vicinity of the visually identified statistically significant peak. The 
reduced $\chi^{2}$ fits range between 1 and 2 and the probabilities of a 
good fitting are always $\geq 0.2$. The results are presented in column 9 of 
Table \ref{tab2}. For spectroscopically confirmed systems, despite the small 
number statistics, there is fair agreement between the spectroscopic and 
photometric cluster redshift estimates. The only exception is object \#16 in 
Table \ref{tab2} whose photometric distribution leads to a peak at 
$z_{\rm p}\sim 0.36$, while $z_{\rm s}=0.782$ well beyond the point after 
which our $z_{\rm p}$ estimates are reliable (cf. section \ref{opti}). 

\begin{figure*} 
\mbox{\epsfxsize=8.0cm \epsffile{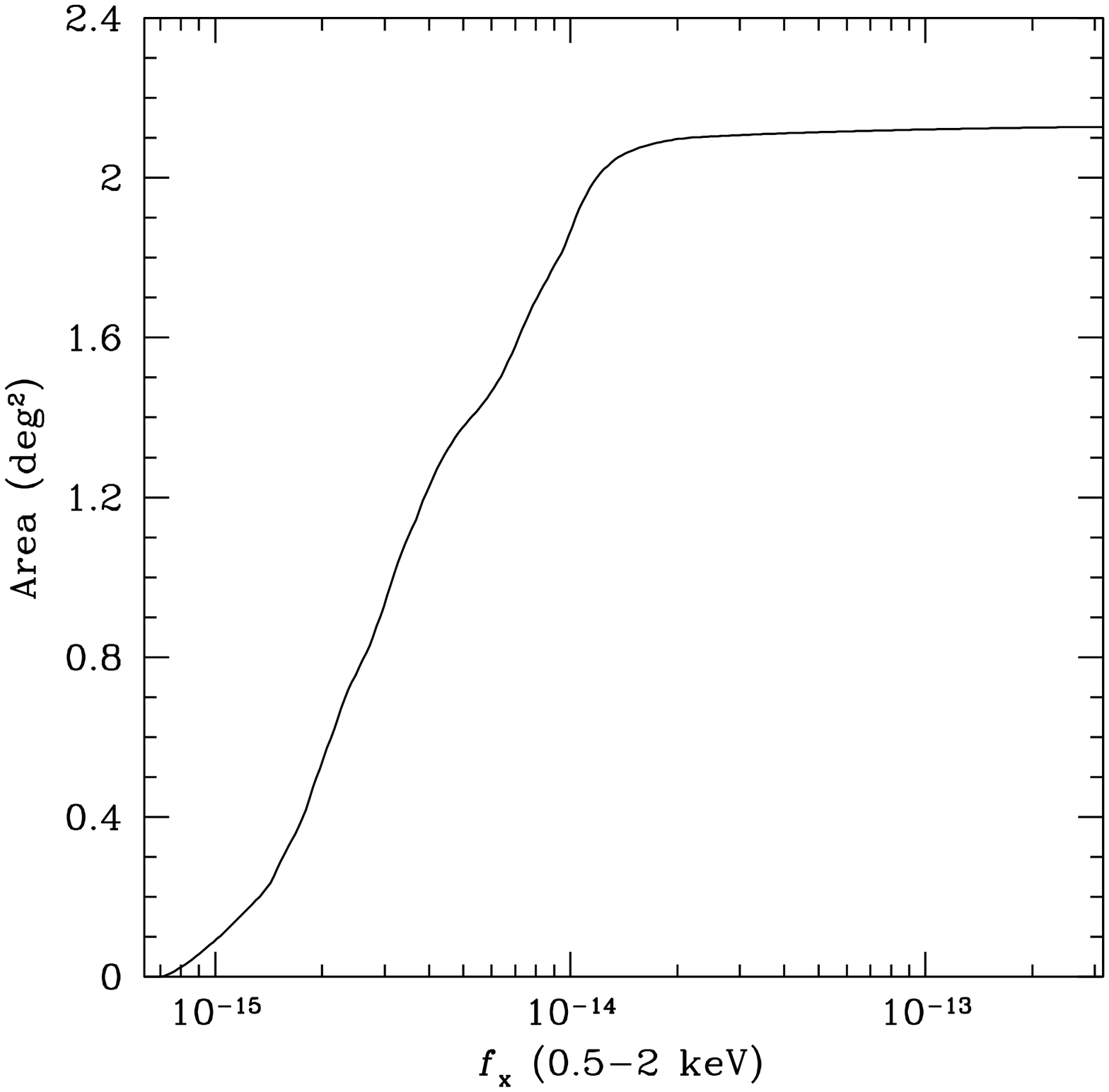}}
\mbox{\epsfxsize=8.0cm \epsffile{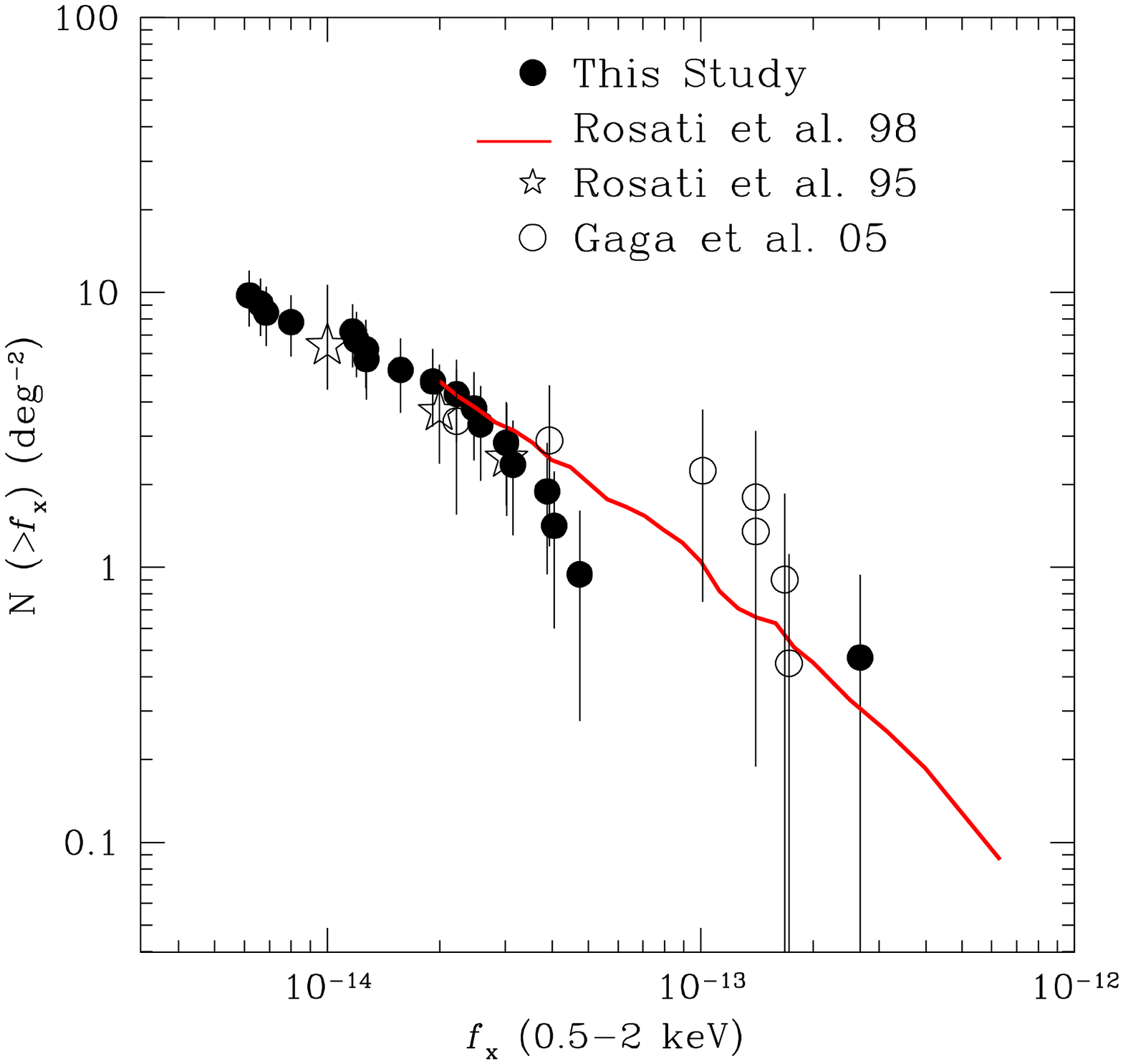}}
\caption{{\bf Left panel}: Area curve for extended sources in our survey 
(see text for details). {\bf Right panel}: Observed cluster cumulative number 
counts for the 19 serendipitous X-ray selected clusters in this paper 
(filled circles) compared to the cluster $\log N - \log S$ derived by Rosati 
et al. (1995; 1998) and that by GPB05. The errorbars are Poissonian estimates.}
\label{area} 
\end{figure*}

\section{The cluster {\boldmath $\log N - \log S$}}

Using the cluster candidates presented in this paper we construct the
$\log N - \log S$ for X-ray selected clusters. The survey sky coverage for 
extended sources is estimated by assuming a mean cluster size of 
$\sim 13^{''}$ typical of the extent of our sources. A circular aperture with 
that radius is slided across the survey area to estimate at each position 
the $5\,\sigma$ fluctuations of the 0.5-2\,keV background counts using the 
background maps, generated as a by-product of the source detection. These are 
then divided by the corresponding exposure time from the exposure map and 
converted to flux assuming a Raymond-Smith SED with temperature 
$T\sim 2$\,keV and Galactic absorption appropriate for each field. We finally 
correct these fluxes for the emission outside the aperture used to sum the 
counts adopting a King's surface brightness profile  with $\beta=0.7$ and 
$r_{\rm c}=0.1\,h^{-1}$\,Mpc (cf. GPB05). This procedure closely resembles 
our $5\,\sigma$ cluster detection and flux estimation methods applied to the 
{\it XMM-Newton} fields used in this paper and described in 
section \ref{xray}. 

The area curve measuring the solid angle available to an extended source of 
a given 0.5-2\,keV flux is shown in the left plot of Figure \ref{area}. We 
note that this curve is not particularly sensitive to the choice of the 
aperture size used to sum the background counts, the SED adopted to convert 
count rates to fluxes or the correction factor to total flux. The X-ray 
cluster $\log N - \log S$ using the extended X-ray sources in our sample 
(with the exception of the two target clusters) is plotted in the right panel 
of Figure \ref{area}. The results of Rosati et al. (1995, 1998) using 
${\em ROSAT}\,$ PSPC data and those of GPB05 based on a serendipitous but 
shallower XMM-{\em Netwon} survey are also shown. Our present survey 
nicely complements these studies reaching fainter flux limits. Moreover, 
there is a fair agreement between all the above samples within the 1$\sigma$ 
uncertainties.

\section{Discussion and conclusions}

We use a total of 15 public {\it XMM-Newton} pointings overlapping with 
5-band optical data from the XID programme  to serendipitously identify 
X-ray selected clusters. In this first paper we present the selection of our 
cluster candidates, their optical properties, including their photometric 
and spectroscopic redshifts when available and their surface  density to the 
limit $f_{\rm x}\,(\rm 0.5-2\,keV)\approx 6\times 
10^{-15}\,erg\,s^{-1}\,cm^{-2}$. 

We use the SAS packages to identify X-ray sources with high 
probability ($P > 99.9 \%$) of being extended. After excluding spurious 
detections clearly associated with double X-ray point sources or falling on 
CCD gaps we identify a total of 21 bona-fide X-ray extended sources over a 
$\approx 2.1\,\rm deg^2$ area which we call the SEXCLAS sample. In this 
sample there are 7 spectroscopically confirmed clusters from the literature, 
2 of which are {\it XMM-Newton} targets and hence not serendipitous sources.

We further exploit the optical multiwaveband data available for our fields 
to explore the optical properties of our sources and attempt to constrain 
their redshifts using photometric techniques. Firstly, we use the percolation 
technique described by BPG04 and identify galaxy overdensities in the 
vicinity of X-ray selected clusters for about 2/3 of our sources. This 
fraction is in fair agreement with previous studies on the optical properties 
of X-ray clusters using data reaching  depths similar to those employed here 
($r\approx 23$\,mag; BPG04; Plionis et al. 2005). The sources that are not 
linked with optical galaxy overdensities are either high-$z$ clusters 
requiring deeper observations to be identified at optical wavelengths, or 
groups with too few members to stand out against the background/foreground 
galaxy surface density.

Next we attempt to use photometric techniques to estimate the redshifts of 
our cluster candidates. We construct the $z_{\rm p}$ distribution of all 
galaxies in the vicinity of X-ray selected clusters and compare it with 
that of all optically selected galaxies from all 15 {\it XMM-Newton} 
pointings. Using the Kolmogorov-Smirnof statistical test we identify peaks 
in these distributions likely to be associated with the cluster candidates. 
For the spectroscopically confirmed clusters the agreement between 
$z_{\rm p}$ and $z_{\rm s}$ is  good. We note however, that our method is 
insensitive to high-$z$ clusters ($z > 0.5$) that lie beyond the magnitude 
limit of the existing optical data (Schuecker et al. 2004).  

Finally, using our sample of 19 serendipitous cluster candidates (e.g. after 
excluding the two targets) we construct the {\it XMM-Newton} cluster 
$\log N - \log S$ to the limit $f_{\rm x}\,(\rm 0.5-2\,keV)\approx 6\times 
10^{-15}\,erg\,s^{-1}\,cm^{-2}$. The estimated surface density is in fair 
agreement with previous {\em ROSAT} and recent {\it XMM-Newton} results. We 
note however, that the faint end in the right panel of Figure \ref{area} 
might be affected by incompleteness due to high-$z$ clusters with poor 
signal--to--noise ratio that does not allow reliable classification of 
their X-ray morphology (e.g. extended). There is at least one such example 
in our survey: a cluster at $z\approx 1.3$ that is paired with an X-ray 
source, albeit a point-like one. Deeper X-ray data are therefore demanded to 
probe thoroughly the $\log N - \log S$ for fluxes 
$\la 5\times 10^{-15}\rm\,erg\,s^{-1}\,cm^{-2}$.    

\section*{Acknowledgments}
This work is funded by the European Union and the Greek Ministry of 
Development in the framework of the Programme 'Competitiveness-Promotion of 
Excellence in Technological Development and Research-Action 3.3.1', Project 
'X-ray Astrophysics with ESA's mission XMM', MIS-64564.

\end{document}